\documentclass[default]{sn-jnl}
\usepackage{etoolbox}
\usepackage{setspace}
\usepackage{bm}
\usepackage{amssymb}
\usepackage{braket}
\usepackage{placeins}
\usepackage{lineno}
\usepackage{bm}
\usepackage{amssymb}
\usepackage{braket}
\usepackage{placeins}
\usepackage{graphicx}%
\usepackage{multirow}%
\usepackage{amsmath,amssymb,amsfonts}%
\usepackage{amsthm}%
\usepackage{mathrsfs}%
\usepackage[title]{appendix}%
\usepackage{xcolor}%
\usepackage{textcomp}%
\usepackage{manyfoot}%
\usepackage{booktabs}%
\usepackage{algorithm}%
\usepackage{algorithmicx}%
\usepackage{algpseudocode}%
\usepackage{listings}%
\usepackage{commath}
\usepackage{natbib}
\setcitestyle{round}
\makeatletter
\renewcommand{\@biblabel}[1]{#1.}
\makeatother
\usepackage{etoolbox}
\makeatletter
\patchcmd{\ps@headings}
{\hbox to \hsize{\hfill Springer Nature 2021 \LaTeX\ template\hfill}}
{\hbox to \hsize{\hfill  \hfill}}
{}
{}
\patchcmd{\ps@headings}
{\hbox to \hsize{\hfill Springer Nature 2021 \LaTeX\ template\hfill}}
{\hbox to \hsize{\hfill  \hfill}}
{}
{}
\patchcmd{\ps@titlepage}
{\hbox to \hsize{\hfill Springer Nature 2021 \LaTeX\ template\hfill}}
{\hbox to \hsize{\hfill  \hfill}}
{}
{}
\makeatother
\jyear{2023}%

\theoremstyle{thmstyleone}%
\theoremstyle{thmstyletwo}%

\theoremstyle{thmstylethree}%
\def\um{\,\textmu m}
\def\nm{\,nm}
\begin{document}

\title[]{Deep and Dynamic Metabolic and Structural Imaging in Living Tissues}
%\subtitle{\small Deep and Dynamic Metabolic and Structural Imaging}
\author[1,2]{\large Kunzan Liu}
\author[1,2]{\large Honghao Cao}
\author[1,2]{\large Kasey Shashaty}
\author[1,2]{\large Li-Yu Yu}

\author[3]{\large Sarah Spitz}
\author[3]{\large Francesca Michela Pramotton}
\author[3]{\large Zhengpeng Wan}

\author[3]{\large Ellen L. Kan}
\author[3]{\large Erin N. Tevonian}

\author[4]{\large Manuel Levy}
\author[4]{\large Eva Lendaro}

\author[3,5]{\large Roger D. Kamm}
\author[3,5]{\large Linda G. Griffith}
\author[4]{\large Fan Wang}

\author*[1,2]{\large Tong Qiu}\email{\small qiutong@mit.edu}
\author*[1,2]{\large Sixian You}\email{\small sixian@mit.edu}

\affil[1]{\small \centering{Research Laboratory of Electronics,}\\Massachusetts Institute of Technology\vspace{1mm}}
\affil[2]{\small \centering{Department of Electrical Engineering and Computer Science,}\\Massachusetts Institute of Technology\vspace{1mm}}
\affil[3]{\small \centering{Department of Biological Engineering,}\\Massachusetts Institute of Technology\vspace{1mm}}
\affil[4]{\small \centering{Department of Brain and Cognitive Sciences,}\\Massachusetts Institute of Technology\vspace{1mm}}
\affil[5]{\small \centering{Department of Mechanical Engineering,}\\Massachusetts Institute of Technology\vspace{1mm}}

\abstract{
Label-free imaging through two-photon autofluorescence (2PAF) of NAD(P)H allows for non-destructive and high-resolution visualization of cellular activities in living systems. However, its application to thick tissues and organoids has been restricted by its limited penetration depth within 300\um, largely due to tissue scattering at the typical excitation wavelength ($\sim$750\nm) required for NAD(P)H. Here, we demonstrate that the imaging depth for NAD(P)H can be extended to over 700\,\textmu m in living engineered human multicellular microtissues by adopting multimode fiber (MMF)-based low-repetition-rate high-peak-power three-photon (3P) excitation of NAD(P)H at 1100\,nm.
This is achieved by having over 0.5\,MW peak power at the band of 1100$\pm$25\,nm through adaptively modulating multimodal nonlinear pulse propagation with a compact fiber shaper. Moreover, the 8-fold increase in pulse energy at 1100\,nm enables faster imaging of monocyte behaviors in the living multicellular models. These results represent a significant advance for deep and dynamic metabolic and structural imaging of intact living biosystems. The modular design (MMF with a slip-on fiber shaper) is anticipated to allow wide adoption of this methodology for demanding \textit{in vivo} and \textit{in vitro} imaging applications, including cancer research, autoimmune diseases, and tissue engineering.
}

\maketitle

%\textbf{Teaser: }720-\textmu m deep NAD(P)H redox imaging of intact living human multicellular microtissues enabled by accessible multimode fiber sources.

\section*{Introduction}\label{sec:main}

Capturing the metabolic dynamics of intact and living biosystems is essential in biomedicine from fundamental research to clinical pathology.
Label-free two-photon autofluorescence (2PAF) imaging of NAD(P)H and FAD enables non-destructive, high-resolution, and three-dimensional (3D) visualization and characterization of cellular metabolic activities. The use of two-photon (2P) excitation of NAD(P)H imaging has been adopted in major tissue types and diseases for non-invasive assessment of oxidative phosphorylation and glucose catabolism in living cells \cite{georgakoudi2002nad,kolenc2019evaluating,liu2018mapping,skala2007vivo,walsh2013optical,gordon2008brain,heaster2021intravital,walsh2021classification}. %[quoted from and Ref: Kyle quinn].

However, NAD(P)H imaging rarely extends beyond 300\um~due to light scattering and out-of-focus background  \cite{kolenc2019evaluating,skala2007vivo,you2018intravital}. Compared to the widely adopted 2P-based labeled imaging \cite{helmchen2005deep}, this problem of limited depth penetration in NAD(P)H imaging is exacerbated by two intrinsic properties of NAD(P)H autofluorophore:
(1) 2P cross-section of NAD(P)H is approximately 4 orders of magnitude lower than the commonly used labels such as green fluorescent proteins (GFPs) ($10^{-2}$\,GM \cite{huang2002two} compared to $10^{2}$\,GM of Calcium Crimson with Ca$^{2+}$ \cite{xu1996multiphoton});
(2) the excited electronic states of NAD(P)H require more blue-shifted excitation wavelengths compared to existing markers ($\sim$750\,nm for NAD(P)H \cite{huang2002two,zipfel2003live} compared to $\sim$930\,nm for GFP \cite{xu1996multiphoton}), leading to increased scattering in deep tissue.

In this work, we demonstrate the depth limit of NAD(P)H imaging can be significantly alleviated by adopting three-photon (3P) excitation of NAD(P)H at 1100\,nm from a multimode fiber (MMF) source. In label-based imaging, it has been proposed and demonstrated that 3P excitation of fluorescent markers can significantly reduce scattering and enhance signal-to-background ratio (SBR) in deeper tissue \cite{horton2013vivo,wang2020three,choe2022intravital,wang2018three,wang2024supercontinuum,wang2023gentle}. Nevertheless, this has not been successfully extended to label-free NAD(P)H imaging. One pragmatic technical challenge is the lack of existing commercial femtosecond sources near the band of 1100\,nm with high peak power and decent beam quality. This work demonstrates the feasibility of using easily accessible MMFs for high-quality deep and dynamic metabolic and structural imaging of living biosystems. 

Prior work, abbreviated as SLAM (simultaneous label-free autofluorescence-multiharmonic) imaging, demonstrated the possibility of using photonic crystal fiber (PCF)-based 1110-nm sources for 3P excitation of NAD(P)H \cite{you2018intravital}. The introduction of 1110-nm excitation enabled simultaneous acquisition of third harmonic generation (THG) originating at the structural interfaces \cite{weigelin2016third,debarre2006imaging,yildirim2022label}, together with the 3P excitation of NAD(P)H and 2P excitation of FAD, plus second harmonic generation (SHG) from collagen fibers \cite{campagnola2003second,zoumi2002imaging}. The combination of these orthogonal contrasts allowed simultaneous visualization of diverse cellular and extracellular components in the unperturbed tissue microenvironment \cite{you2018slide,you2019label,you2019real,you2021label,park2023vivo}. Nevertheless, it was not clear how deep SLAM can image into living tissues (the depth was limited to 200\,\textmu m due to the energy constraint of excitation pulses) and how widely deployable SLAM could be for the general bioimaging field given the non-trivial PCF and pulse shaper setup. 

In this work, we demonstrate for the first time that the depth limit of NAD(P)H imaging can be extended to over 700\,\textmu m, using living engineered human multicellular microtissues as test samples, and by adopting MMF-based low-repetition-rate high-peak-power 3P excitation of NAD(P)H at 1100\,nm, enabling deep and dynamic SLAM (dSLAM) imaging. The high peak power exceeding 0.5\,MW at the band of 1100$\pm$25\,nm was obtained by adaptively modulating multimodal nonlinear pulse propagation with a compact fiber shaper \cite{qiu2024spectral}. Furthermore, the 8-fold increase of pulse energy at 1100\,nm allows us to capture faster monocyte behaviors in the engineered human multicellular microtissues \textit{in vitro}. These results and findings represent an important advance towards deeper and faster metabolic and structural imaging of intact living biosystems. We anticipate the flexibility provided by the modular design (step-index (SI) MMF with a slip-on fiber shaper) will allow the proposed imaging methodology to be widely adopted for demanding \textit{in vivo} and \textit{in vitro} imaging applications \cite{xu2013recent}, including cancer research \cite{entenberg2023intravital,walsh2014quantitative,friedl2003tumour,condeelis2006macrophages}, autoimmune diseases \cite{pham2025perspectives,friedl2004prespecification}, and tissue engineering \cite{zervantonakis2012three,griffith2006capturing}.
% [Ref redox applications, organoid imaging etc, cell painting drug screening etc]. 

\section*{Results}\label{sec:results}

\subsection*{SI MMF for high-quality high-peak-power metabolic and structural imaging}
\label{tuning}

\begin{figure}[t]
\centering
\includegraphics[width=1\textwidth]{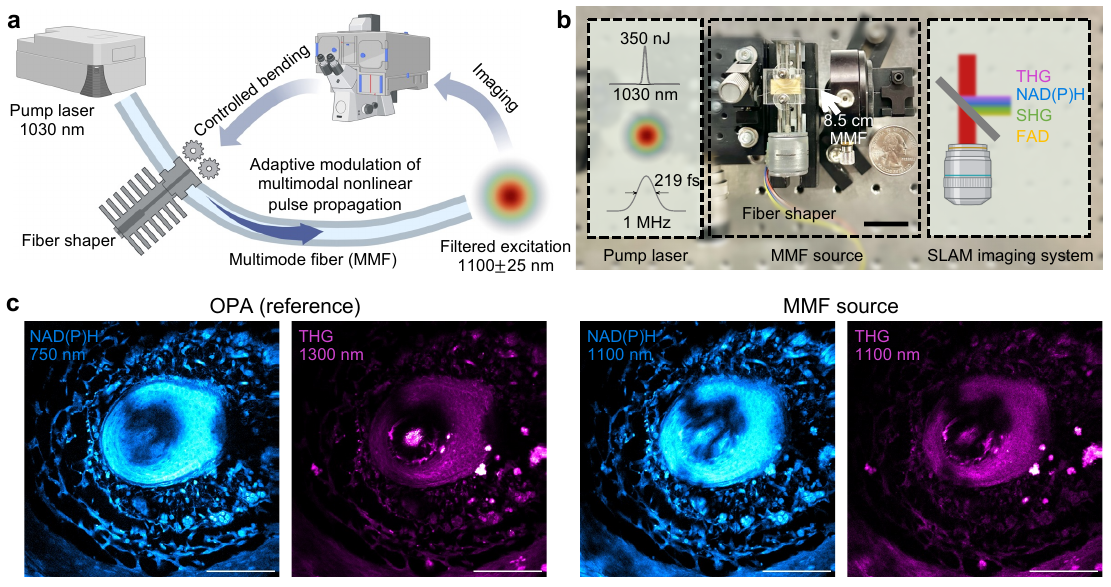}
% \label{fig1}
\caption{\textbf{SI MMF as a compact, accessible, and high-quality label-free imaging source.} \textbf{a}, Schematic for working principles. High-energy ultrashort pulses from pump laser (up to 350\,nJ, 219\,fs) at 1030\,nm were injected into the SI MMF, where the propagation of multimodal nonlinear pulses were adaptively modulated by a compact fiber shaper that applies controlled macro-bending. The spectrally-filtered fiber output at 1100\,nm is directed to the microscopy system for label-free metabolic and structural imaging, where the imaging signal is used as the feedback for controlling the fiber shaper.
\textbf{b}, Photograph (middle panel) of the MMF source based on a compact slip-on fiber shaper and its comparison with a US quarter, highlighting the compactness and the modular nature of the MMF source. Left and right panels: illustrative diagrams of the input pulses and the imaging system. Scale bar: 30\,mm. \textbf{c}, Comparison between images acquired with a commercial OPA and with the MMF source at the same site of the mouse whisker pad tissue. The OPA images were captured sequentially at 750\,nm and 1300\,nm, whereas the MMF source images were captured simultaneously at 1100\,nm. Scale bars: 100\,\textmu m. OPA: optical parametric amplifier.}
\label{fig1}
\end{figure}

To create a high-peak-power and accessible fiber source at the band of 1100\,nm, we chose (1) a standard silica-core SI MMF as the medium for its relatively large mode area and power scalability \cite{wright2022nonlinear,krupa2019multimode,wright2022physics,wright2017spatiotemporal,wei2020harnessing,ding2021spatiotemporal,ji2023mode,krupa2017spatial}, and (2) a slip-on compact fiber shaper as the control device for its low cost and ease of use \cite{qiu2024spectral} (Fig.~\ref{fig1}a, b). 
% In short, these choices afford imaging quality comparable to a commercial laser source operating at different wavelengths (Fig.~\ref{fig1}c).
% Next, we will further detail the underlying mechanisms behind this capability. 
Ultrashort pulses (Light Conversion, Carbide) at 1030\,nm of 1\,MHz repetition rate were launched into a 8.5-cm long SI MMF (25\,\textmu m core, 0.1\,NA) and the pulse propagation was modulated by controlled bending through a slip-on fiber shaper (see Methods). The input peak power up to 1.60\,MW (350\,nJ pulse energy) is well below the critical power of silica at around 4.7\,MW \cite{dubietis2017ultrafast}. 

\begin{figure}[t]
\centering
\includegraphics[width=\textwidth]{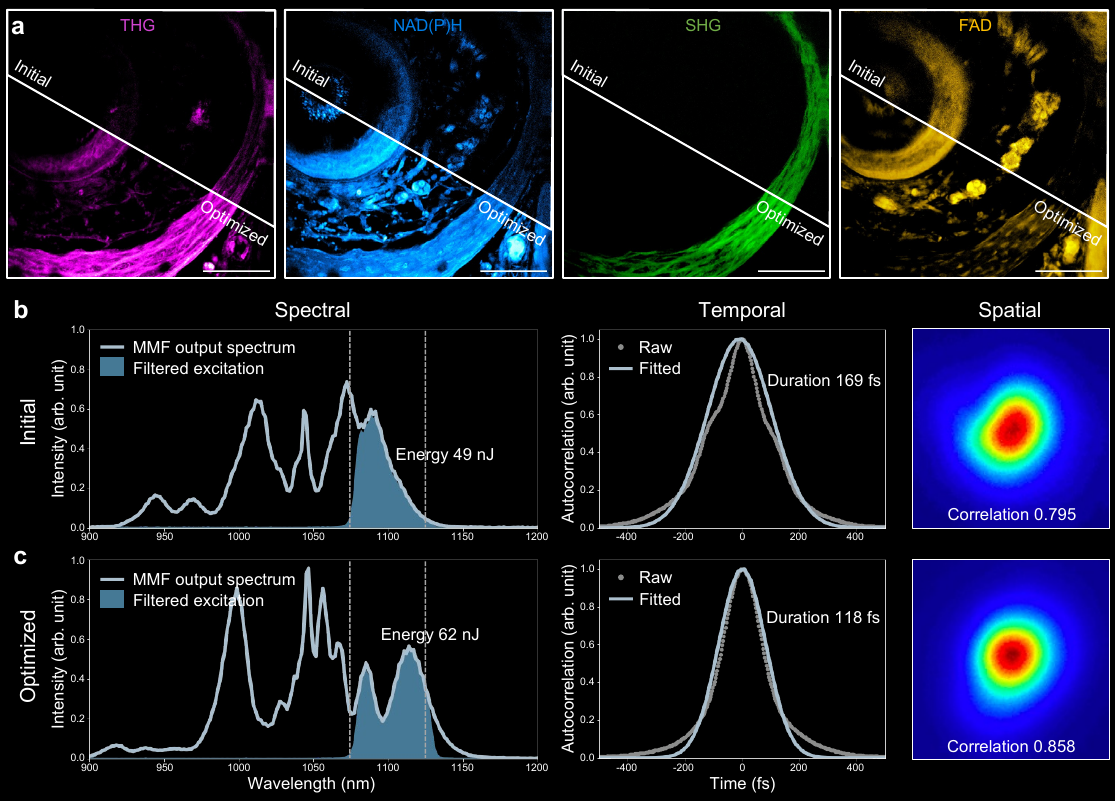}
\caption{\textbf{Shaping the SI MMF to optimize imaging SNR and resolution.} \textbf{a}, Comparison of images at the same site of the mouse whisker pad tissue acquired before (initial) and after (optimized) optimizing the fiber shaper, with the same contrast adjustments. Scale bars: 100\,\textmu m. Spectral, temporal, and spatial characterizations of the fiber output at the spectral band of $1100\pm25$\,nm before (\textbf{b}) and after (\textbf{c}) optimizing the fiber shaper. The spectra before filtering are also presented (solid blue lines in the left panel). Duration is estimated based on the autocorrelation signals and with the assumption of Gaussian pulse shape. Spatial profiles represent the near-field intensity distributions of the fiber output; numbers denote the correlation with the fundamental LP$_{01}$ mode.}
\label{fig2}
\end{figure}

As expected, the higher nonlinearity associated with 3P processes made THG and NAD(P)H imaging highly susceptible to the deteriorating beam quality and temporal duration that are known to be associated with the MMF output field \cite{xu2002multiphoton}. Building on the idea of modulating multimodal nonlinear pulse propagation in the recent work \cite{qiu2024spectral}, we developed a compact version of the fiber shaper for an off-the-shelf 8.5-cm few-mode fiber (25\,\textmu m core size) to address the need of better beam quality and shorter temporal duration, to enable deeper and dynamic SLAM imaging (Fig.~\ref{fig1}b). Compared to the previous 3D printed 5-actuator (to 8-actuator) fiber shaper, the relatively shorter pulses are obtained by using a shorter fiber (30\,cm vs 8.5\,cm) and a compact laser-cutting-based fiber shaper (consisting of one downsized actuator, see more in Methods). The near-Gaussian beam was obtained by modulating the multimodal pulse propagation within a few-mode fiber (25\,\textmu m core size) and spectral filtering of the red-shifted wavelengths at 1100\,nm. Similar to the previously demonstrated fiber shaper, changing the position of the actuator applies controlled macro-bending on the fiber, which produces local refractive index perturbations and alters the propagation of the nonlinear multimodal pulses, and thus customizes the MMF output in the spectral-temporal-spatial properties. By integrating the signal level of THG imaging as the real-time feedback, the actuator position can be actively adjusted to produce a spectral-temporal-spatial profile that is optimal for 3P imaging (Fig.~\ref{fig2}). 

To investigate the quality of the images that can be obtained with our MMF source, we compare the images out of the optimized MMF source with a state-of-the-art optical parametric amplifier (OPA) (Light Conversion, Cronus-3P) (NAD(P)H/THG using 750\,nm/1300\,nm from OPA and 1100\,nm from MMF) at the same site of the mouse whisker pad tissue. The results presented in Fig. \ref{fig1}c show that the signal-to-noise ratio (SNR) and the resolution are comparable between the two image sets. The NAD(P)H image out of the MMF source looks even more highly resolved than that of the OPA, but this is very likely to be the consequence of 3P excitation rather than a result of the difference in beam quality (see more in the next Result section). 

Since this is an end-to-end adaptive optimization, characterization of the MMF output field in the spectral, temporal, and spatial domains was performed during imaging to provide more information into the mechanism of the improvement. 
A representative illustration of this optimization process is presented in Fig. \ref{fig2}, where it shows that the signal enhancement of the multiphoton processes can be reliably achieved by adaptively adjusting the position of the single-actuator fiber shaper guided by the signal level of THG imaging. 
%The fiber shaper is capable of identifying a position that results in increased pulse energy, reduced pulse duration, and/or a more Gaussian-like spatial profile within the targeted spectral band of 1100 nm (Fig. \ref{fig2}b, c). 
Specifically in this example of imaging the mouse whisker pad tissue, from the initial to the optimized fiber shape, a 1.27-fold increase in band energy and a 1.43-fold reduction in pulse duration, with an improvement of beam spatial profile were obtained (Fig. \ref{fig2}b, c; see Methods for correlation calculation). The theoretical signal enhancement for 3P (4.16-fold) and 2P (2.29-fold), calculated using the enhancement in band energy and reduction in pulse duration \cite{hontani2021multicolor}, are comparable to the measured SNR enhancements from Fig.~\ref{fig2}a, 5.22-fold in 3P processes and 2.70-fold in 2P processes (see Methods for SNR calculation). The slight discrepancy is likely due to the simple assumptions of Gaussian pulse shape and Gaussian beam shape. These results show that, by using the compact fiber shaper on the few-mode SI fiber, a 120-fs near-Gaussian fiber output can be reliably obtained for high-quality high-peak-power metabolic (3P-NADH) and structural (THG) imaging. In addition, the entire wavelength conversion unit is highly compact, modularized, and low-cost (Fig. \ref{fig1}b). These results demonstrate the feasibility of using a standard SI MMF with a compact fiber shaper for accessible, high-quality, and high-peak-power excitation at 1100\,nm. 

\subsection*{Feasibility of using 1100\,nm for deep 3PAF NAD(P)H imaging}
\label{3P}

\begin{figure}[t]
\centering
\includegraphics[width=\textwidth]{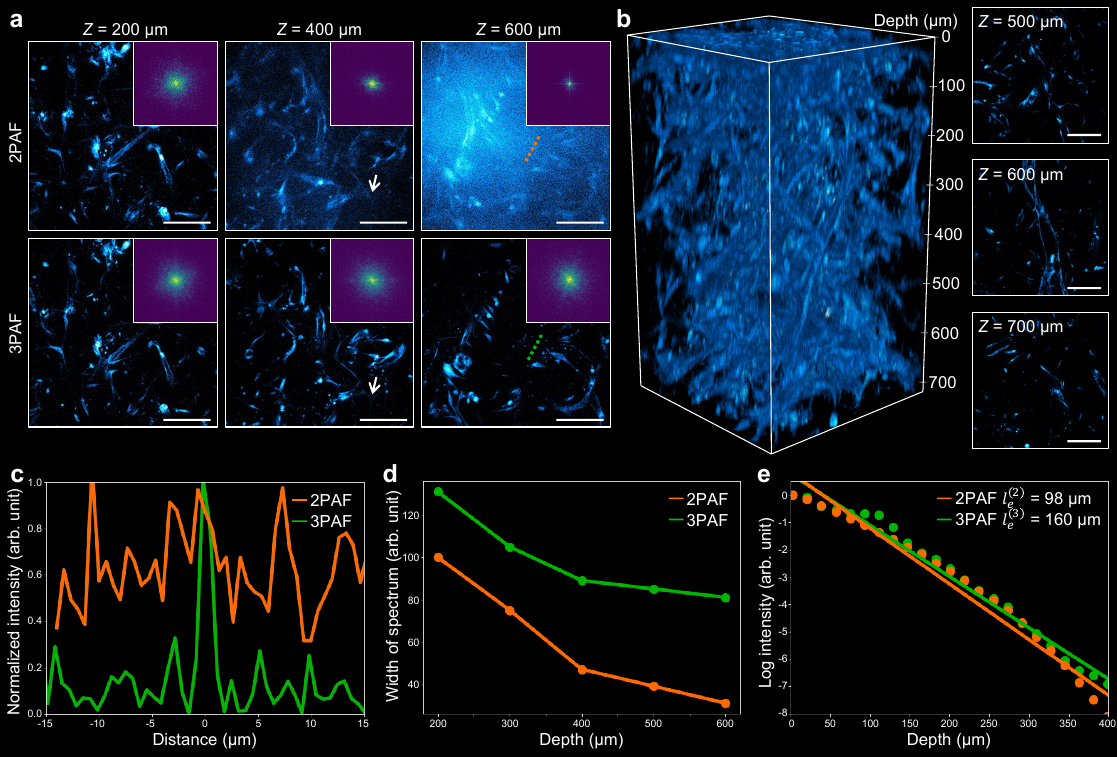}
\caption{\textbf{Deep NAD(P)H imaging with 1100\,nm MMF source.} 
\textbf{a}, Comparison of 2PAF and 3PAF NAD(P)H imaging of intact 3D microvascular network at different depths, applying identical contrast adjustments for the 2PAF and 3PAF images at the same depth. Magnitude of the Fourier transform of individual images are plotted in the insets, indicating differences in the resolution of the images.
\textbf{b}, 3PAF NAD(P)H imaging of the 3D microvascular network through the entire 720\,\textmu m of depth, with 2D images shown at different imaging depths.
\textbf{c}, Normalized intensity profiles along the co-registered lines in \textbf{a}.
\textbf{d}, Width of the magnitude of the Fourier transform across different depth in 2PAF and 3PAF NAD(P)H imaging.
\textbf{e}, Signal intensity of 2PAF and 3PAF as a function of imaging depth measured in the same site of \textbf{b}. The effective attenuation lengths (EALs) $\ell_e^{(2)}$ and $\ell_e^{(3)}$ were fitted for 2PAF and 3PAF, respectively.
Scale bars: 100\,\textmu m.}
\label{fig3}
\end{figure}

To investigate the depth limitation of NAD(P)H imaging and the effectiveness of using 1100\,nm to extend the depth limit, we performed NAD(P)H autofluorescence imaging with 2P excitation at 750\,nm and 3P excitation at 1100\,nm. For comparison, the 2PAF NAD(P)H imaging was performed using the signal path from the OPA at 750\,nm, and the 3PAF NAD(P)H imaging at the same site was performed using the MMF-based 1100\,nm source using a 3D microvascular network \cite{zhang2021vascularized,wan2022robust}, with varying pulse energy according to the imaging depths that maintains $\sim$3\,nJ pulse energy at the excitation focus. The microvascular network is an engineered human multicellular microtissue that consists of dense and complex 3D network formed by vascular endothelial cells, which was chosen as test samples since the vascular endothelial cells exhibited stable NAD(P)H signals through the entire 720\,\textmu m of depth (Fig. \ref{fig3}b). 

In superficial layers ($Z\leq 200$\,\textmu m), the NAD(P)H images generated using 2PAF have comparable SNR and resolution as the 3PAF using 1100\,nm. However, as the imaging delves deeper, due to the out-of-focus background signal, the SBR degrades much faster in 2PAF (Fig. \ref{fig3}a; see Methods for SBR calculation).
For example, at the depth of $Z=600$\,\textmu m, the SBR of 2PAF dropped to 1.12, compared to 6.68 for 3PAF. The low SBR can lead to failure in identifying cells (white arrow in Fig. \ref{fig3}a at $Z=400$\,\textmu m) and structures (lines in Fig. \ref{fig3}a at $Z=600$\,\textmu m, with intensity profiles in Fig. \ref{fig3}c).

Next, we quantified the image degradation and signal attenuation of both 2P and 3P nonlinear processes in relation to imaging depth in Fig. \ref{fig3}d and \ref{fig3}e. We observed that a significant contributor to the degradation of the 2PAF NAD(P)H imaging is the overwhelming background and the blurred high-frequency details in the spatial domain of the 2D image slice, which corresponds to a significantly increased dominance of the low-frequency zone and decreased coverage of the high-frequency zone in the frequency domain, as visualized by the spectrum of the Fourier transform shown in the insets of Fig. \ref{fig3}a. Qualitatively, we computed and plotted the width of the spectrum of the Fourier transform in Fig. \ref{fig3}d across different depths. As the imaging depth increases, the resolution degradation occurs much faster in the 750\,nm 2P excitation scheme.
In addition, we characterized the effective attenuation lengths (EALs) of NAD(P)H, defined by the depth at which the fluorescence signal attenuates by $1/e^{2}$ and $1/e^{3}$ for 2P and 3P imaging, respectively \cite{choe2022intravital}. The EAL of 1100\,nm 3PAF NAD(P)H imaging was measured as $\ell_{e}^{(3)}=160$\,\textmu m, 1.63 times longer than that of $\ell_{e}^{(2)}=98$\,\textmu m in 750\,nm 2PAF (Fig. \ref{fig3}e), which benefits from the use of longer excitation wavelength in 3P imaging \cite{horton2013vivo,wang2018three,ouzounov2017vivo,wang2020three}.
% [2013 Chris Xu NP, 2018 Tianyu Wang  Chris Xu NM, 2017 Chris Xu NM, 2020 Tianyu Wang  Chris Xu Optica]
%Consistent with other works in deep labeled 3P mouse brain imaging , longer EAL of 3P allows more efficient energy delivery deep in samples due to reduced light scattering.
These investigations suggest that NAD(P)H imaging in deep tissue requires 1100\,nm high-peak-power 3P excitation for high-resolution and high-contrast imaging throughout the entire depth of the 3D microvascular network, underscoring the potential of dSLAM for deep and dynamic metabolic and structural imaging. 

\subsection*{Deep metabolic and structural imaging of living blood-brain barrier microfluidic model}
\label{deep}

\begin{figure}[htbp]
\centering
\includegraphics[width=\textwidth]{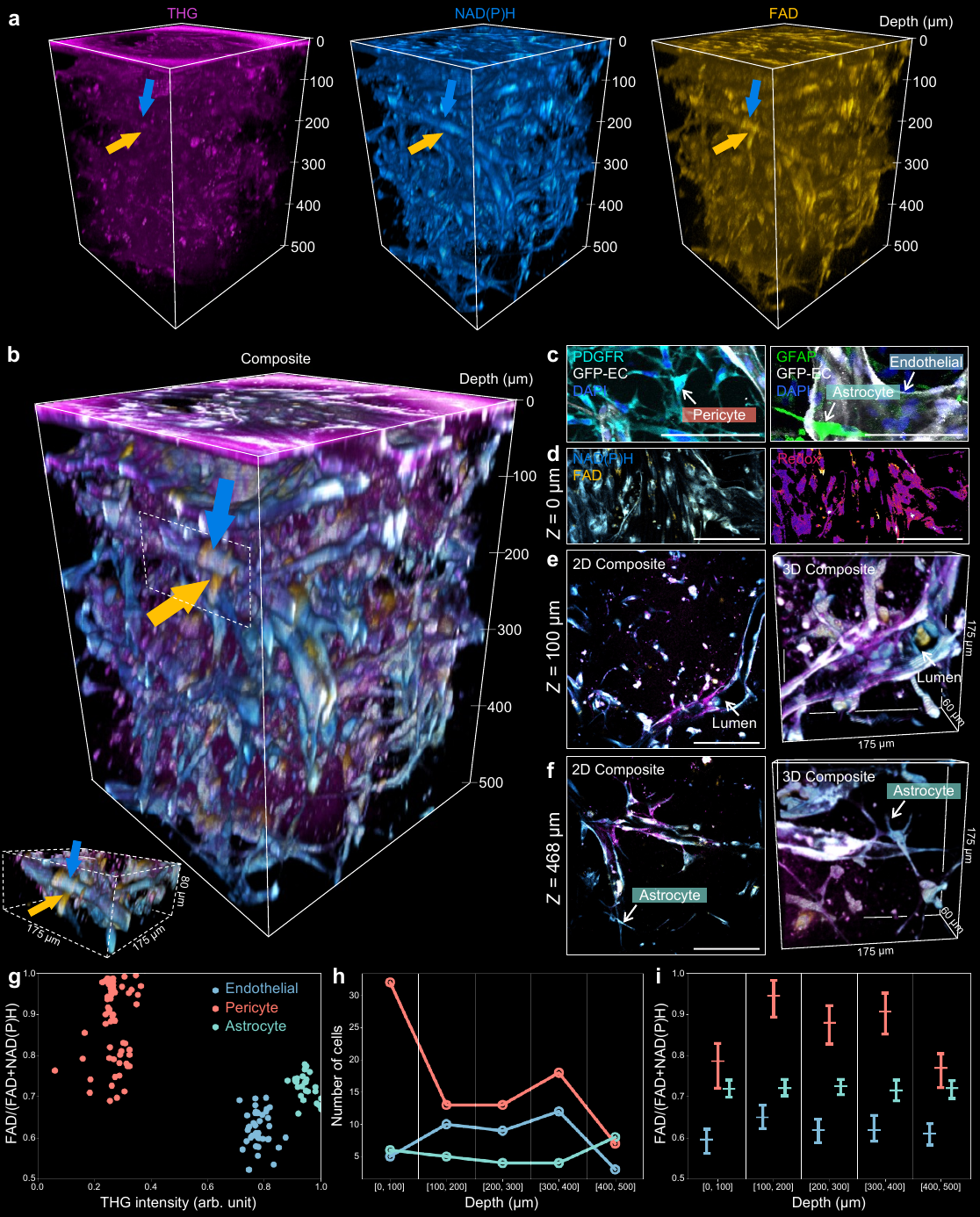}
\caption{\textbf{dSLAM for deep metabolic and structural imaging.}
3D visualization of THG, NAD(P)H, and FAD signals (\textbf{a}) and the merged signals (\textbf{b}) for structural and metabolic imaging of the entire 500-\textmu m-deep living blood-brain barrier microfluidic model, comprising vascular endothelial cells, pericytes, and astrocytes. Dashed volume in \textbf{b}: a FAD-strong pericyte (yellow arrow) wraps an NAD(P)H-strong vascular endothelial cell (blue arrow).
\textbf{c}, Immunofluorescence (IF) imaging of stained vascular endothelial cells, pericytes, and astrocytes.
\textbf{d}, NAD(P)H and FAD imaging and the resulting map of redox ratio in sub-cellular resolution at the shallowest layer (see Methods for redox calculation).
\textbf{e, f}, 2D and 3D visualization at different deeper layers, showing the lumen structure and astrocytes extending projections around vascular endothelial cells in the living blood-brain barrier microfluidic model.
\textbf{g}, Redox ratio and THG intensity of cells in the volume. Cells cluster into the same type based on their structural and metabolic features.
Density (\textbf{h}) and redox ratio (\textbf{i}), of different cell types at different depths of the living blood-brain barrier microfluidic model.
Scale bars: 100\,\textmu m.
}
\label{fig4}
\end{figure}

A major motivation for developing the dSLAM imaging platform is to perform non-invasive deep metabolic and structural characterization of living multicellular systems at the sub-cellular level. To examine the capability of dSLAM for this task, we acquired a 350\,\textmu m$\times$350\,\textmu m$\times$500\,\textmu m image stack of living blood-brain barrier microfluidic models (the entire depth of the model) \cite{hajal2022engineered}, as shown in Fig.~\ref{fig4}. Developing and monitoring blood-brain barrier microfluidic models are essential for improving targeted therapies for neurological disorders \cite{hajal2022engineered}. Deep label-free metabolic and structural imaging of human blood-brain barrier models \textit{in vitro} could aid the development of therapeutics with improved brain delivery. Here we show that, by using the MMF-based 1100-nm excitation with over 0.5\,MW peak power, the deep NAD(P)H imaging integrated with the simultaneously acquired THG and FAD imaging allows 3D redox mapping of the entire intact living human blood-brain barrier model (Fig. \ref{fig4}a, b, see Supplementary Video 1 and 2).

The human blood-brain barrier microfluidic models were developed with self-assembled vascular endothelial cells, pericytes, and astrocytes in a 500-\textmu m-thick microfluidic device (see Methods). 
3D visualization from Fiji \cite{schindelin2012fiji} distinctly shows the three cell types based on their optical signatures in the surrounding fibrin gel (Fig. \ref{fig4}a, b, compared with immunofluorescence (IF) imaging of stained cells in Fig. \ref{fig4}c;  see Methods). For example, vascular endothelial cells, which are engineered to form a perfusable vasculature (blue arrow in Fig. \ref{fig4}b), exhibit high NAD(P)H, consistent with demonstrations that endothelial cells use glycolysis for ATP production \cite{georgakoudi2012optical,pochechuev2019stain,quintero2006mitochondria,falkenberg2019metabolic,you2021label}. Pericytes, wrapping around the vasculature (yellow arrow in Fig. \ref{fig4}b), are observed to have a high concentration of FAD, likely due to their contractile functions \cite{peppiatt2006bidirectional,hall2014capillary,fernandez2010pericytes}. Astrocytes, responsible for regulating ion and neurotransmitter concentrations, exhibit strong THG signals, which is in line with previous reports and might be attributed to their complex branched morphology \cite{witte2011label,pochechuev2019stain}. Further quantification of the redox ratio and the THG signal intensity in Fig. \ref{fig4}g shows distinct cell clusters based on their raw optical signatures.
%The 3D imaging in Fig. \ref{fig4}b also clearly reveals the intimate 3D association between cell types, with endothelial cells (blue arrow) being enveloped by the pericytes (yellow arrow) at $Z=120$\,\textmu m deep in the model, which may otherwise be overlooked without deep metabolic and structural imaging. 

Together with the distinct optical signatures, the deep 3D imaging capability enables non-invasive visualization of the complex multicellular interactions, metabolic states, and depth-dependent variations within the living blood-brain barrier microfluidic model. At the shallowest layer ($Z=0$\,\textmu m, closest to the cover glass bottom), the cells displayed flat-out morphology that's similar to 2D cell cultures (Fig. \ref{fig4}d). As the imaging delves deeper, cells start to exhibit three-dimensional features, such as the forming of the lumen at $Z=100$\,\textmu m (Fig. \ref{fig4}e, arrow pointing to the lumen of the vasculature network formed by vascular endothelial cells), the wrapping actions of pericytes over vascular endothelial cells at $Z=120$\,\textmu m (Fig. \ref{fig4}b, dashed volume with zoom-in visualization on the side, with yellow arrow pointing to the pericytes and blue arrow pointing to the vascular endothelial cells), and astrocytes extending projections around vascular endothelial cells at $Z=468$\,\textmu m (Fig. \ref{fig4}f). Furthermore, deep structural and metabolic imaging reveals depth-dependent metabolic distributions in living cells. For example, as shown in  Fig.~\ref{fig4}i, pericytes in the middle of the blood-brain barrier model ($Z$=100--400\,\textmu m) were observed to have higher FAD intensity compared to the outer layers. 
This phenomenon may arise from the proximity of the glass chamber at the shallowest and deepest layer of the blood-brain barrier microfluidic model, which could impact the pericyte contractile functions, or limit the growth of pericyte due to the constrained orientation of the vasculature network.

\subsection*{Dynamic metabolic and structural imaging of monocyte behaviors \textit{in vitro}}
\label{dynamic}

\begin{figure}[htbp]
\centering
\includegraphics[width=\textwidth]{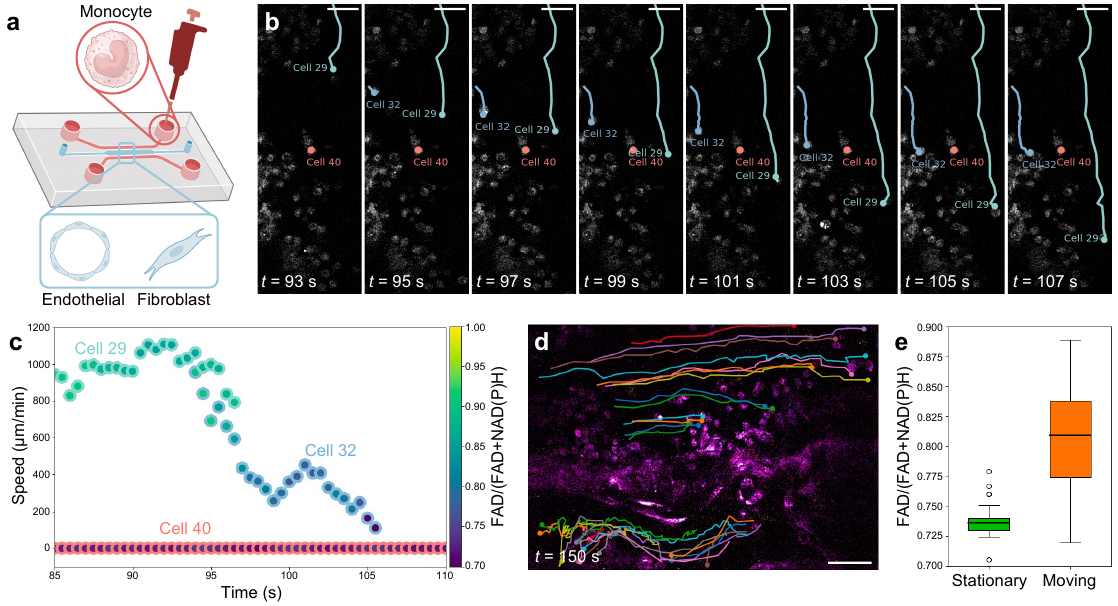}
\caption{\textbf{dSLAM for dynamic metabolic and structural imaging.}
\textbf{a}, Microfludic device setup of \textit{in vitro} imaging of monocyte behaviors in the vasculature network formed by the vascular endothelial cells and fibroblasts.
\textbf{b}, Three motility behaviors of monocytes in the time-lapse imaging, represented by Cell 29, Cell 32, and Cell 40.
\textbf{c}, Speed of the three representative cells during a 25-s time window, with variations in their redox ratio showing the relation between metabolic and motility behaviors.
\textbf{d}, Trajectories of cells analyzed in the entire 150-s time window.
\textbf{e}, The statistics of redox ratio of stationary and moving cells.
Scale bars: 50\,\textmu m.}
\label{fig5}
\end{figure}

Another motivation for developing a label-free metabolic and structural imaging system is to perform non-invasive imaging of cellular dynamics in living and thick biosystems. Previous SLAM imaging demonstrates the capability of SLAM microscopy on tracking the metabolic dynamics of leukocyte migration \textit{in vivo} up to 40\,\textmu m/min \cite{you2018intravital}. The speed was largely limited by the available peak power ($\sim$40\,kW) at the excitation focus. With the peak power of over 0.5\,MW out of filtered excitation of the optimized MMF source, we are able to record time-lapse videos at a pixel rate between 0.5--1\,MHz, which is ultimately limited by the repetition rate of our pump laser.

To examine the capability of dSLAM for capturing faster dynamics, monocyte behaviors were tracked \textit{in vitro} to mimic the immune cell recruitment process in the living vascular network (Fig. \ref{fig5}a, see Methods for sample preparation) \cite{zhang2022interstitial}, with time-lapse imaging at a pixel rate of 0.5\,MHz, which permits a frame rate of 2\,Hz and a field-of-view (FOV) of $0.12\times 0.12 \,\text{mm}^{2}$ for simultaneously monitoring multicellular metabolic and motility behaviors.
In Fig. \ref{fig5}d, the metabolic activities and motility features of 50 cells were characterized over a 150-s imaging session, with captured speeds exceeding 1000\,\textmu m/min (Cell 29 in Fig. \ref{fig5}c). We observed a few motility behavioral patterns during this window (Fig. \ref{fig5}b). As an example, Cell 29 exhibited rapid movement within the lumen and maintained a redox ratio indicative of a high metabolic state, while Cell 40 remained stationary, with a stable redox ratio of $R_{40}=0.73$. Interestingly, Cell 32 showed a marked decrease in both speed and redox ratio before adhering to the vasculature (see Supplementary Video 3). Fig. \ref{fig5}c illustrates the speed and metabolic changes of the three representative cells during the monocyte recruitment. To investigate the potential link between cell recruitment and metabolic activity, Fig. \ref{fig5}e depicts the redox ratio of stationary and moving cells tracked in Fig. \ref{fig5}d, suggesting a lower redox ratio in stationary cells, potentially pointing to increased metabolic activities in monocytes that have attached to the vascular networks. 
The application of this imaging platform to various living tissue models and pathologies could offer distinctive insights into the metabolic changes accompanying immune cell dynamics within multicellular systems.

\section*{Discussion}

The results here demonstrate a significant advancement in the deep and dynamic imaging capabilities of label-free NAD(P)H microscopy in intact living biological systems. By leveraging MMF-based wavelength conversion to generate high-peak-power pulses at 1100$\pm$25 nm band, we were able to extend the practical imaging depth of NAD(P)H to 700\,\textmu m in living engineered human multicellular microtissues. 
This enhancement in imaging depth opens up new possibilities for visualizing metabolic processes and structural features in significantly thicker 3D tissue constructs and small model organisms.
The ability to observe multicellular interactions and metabolic processes at greater depths with improved imaging speed, holds promise for unraveling the complexities of diseases such as cancer, neurodegenerative disorders, and immune-mediated conditions.
Additionally, the utilization of an MMF with a compact slip-on fiber shaper facilitates the delivery of high-peak-power pulses with a near-Gaussian beam profile in a flexible manner, offering potential for miniaturization and widespread adoption compared to prior approaches.

\section*{Materials and Methods} \label{sec:methods}
\subsection*{MMF and beam characterizations}
An SI MMF (Thorlabs, FG025LJA; 25/125 \textmu m, 0.10 NA) of 8.5\,cm length was launched by a high-power femtosecond mode-locked ytterbium laser (Light Conversion, Carbide) at 1030\,nm with a pulse energy of 350\,nJ and a pulse width of 219\,fs. Weak focusing was achieved using an achromatic doublet with 100\,mm focal length (Thorlabs, AC254-100-B) for fiber coupling with a coupling efficiency of 76\%. The output supercontinuum generation pulses were collimated by a 25.4-mm focal length off-axis parabolic mirror (Edmund Optics, 36-586) and spectrally filtered by a 1100$\pm$25\,nm bandpass filter (Edmund Optics, 85-906). A continuously variable neutral density (ND) filter (Thorlabs, NDC-50C-4M-B) was placed before the imaging system to adjust the input pulse energy during imaging. For comparison, the laser reference images in Fig. \ref{fig1}c and Fig. \ref{fig3}a were collected by corresponding wavelengths using OPA (Light Conversion, Cronus-3P). The output pulse spectrum characterization was performed using a NIR spectrometer (Ocean Insight, NIRQuest+1.7). The pulse width was measured using an autocorrelator (Light Conversion, GECO). The near-field output spatial profile was measured using a CMOS-based camera (Mako, G-040B). The metric to characterize the quality of the beam profile in Fig. \ref{fig2} is the correlation of the beam profile and an ideal LP$_\text{01}$ intensity distribution.

\subsection*{Fiber shaper design}
The fiber shaper was built on our previous device \cite{qiu2024spectral} but with a modification to make it smaller. The fiber shaper device was driven by a linear actuator with a 2-phase 4-wire stepper motor, having a motion range of 34\,mm and a step size of 0.025\,mm. A customized fiber holder was fabricated using acrylic by a laser cutter (Universal, PLS6.75) and was assembled on the actuator slider. A microcontroller unit (Arduino, Uno Rev3) was used to control the stepper motor, with codes programmed in Python and sent from a personal computer (PC). To adapt to the short fiber length for minimizing pulse dispersion, a single high-precision actuator was used, with a minimal bending radius of 5\,mm.

\subsection*{Imaging system}
The SLAM microscopy was implemented as an inverted scanning microscope. A galvanometer mirror pair (ScannerMAX, Saturn-5 Galvo and Saturn-9 Galvo) was used to raster scan the 1100\,nm beam out of the MMF. A water immersion objective (Olympus, XLPLN25XWMP2) focused the beam on the imaging plane. A microscope stage (ASI, MS2000) was used to adjust the center of the FOV for mosaic scanning and the focal plane for deep imaging series. The excitation and emission paths were separated by a dichroic mirror (Thorlabs, DMLP650L). The emission signals were further separated into four detection channels using dichroic mirrors (Chroma, T412lpxt, T505lpxr, T570lpxr) and bandpass filters (Chroma, ZET365/20x; Edmund Optics, 84-095; Edmund Optics, 65-159; Semrock, FF01-609/57-25), corresponding to THG, NAD(P)H, SHG, and FAD. Photons were collected by four individual photomultiplier tubes (Hamamatsu, H16201) and signals were translated to images through a custom-written software.
A commercial OPA (Light Conversion, Cronus-3P) emitting pulses at 750\,nm (approximately 40\,fs) and 1300\,nm (46\,fs) was also directed to the same microscope system for acquiring reference images. 

\subsection*{Image processing and analysis}
Raw images were acquired from custom LABVIEW acquisition software. No denoising or deconvolution was applied in the data visualization in this study, in order to present the intrinsic signal level at different depth. The individual images were colored and merged using the Fiji (National Institutes of Health) software \cite{schindelin2012fiji}.
The pseudo-color for the four modalities was chosen as magenta hot for THG, cyan hot for NAD(P)H, green for SHG, and yellow hot for FAD, and were used consistently throughout this study. 
% Custom-software was used for image stitching and stacking.
For Fig. \ref{fig2}a, the imaging SNR was calculated as the average value of the brightest 0.5\% pixels \cite{choe2022intravital}, and the initial and optimized images were displayed with the same contrasts for comparison.
For Fig. \ref{fig3}, the imaging SBR was calculated as the ratio of the average value of the brightest 0.5\% pixels and the average value of the pixels in the generated background mask.
For Fig. \ref{fig5}, the regions of interest (ROI) was selected based on the cells that were moving between each frame. Cell tracing analysis was achieved by locating the center of the moving cells in each frame and plotting the time-lapsed trace, and rolling average was applied to minimize the error in manual labeling.
The optical redox ratio is defined as the ratio of the fluorescence intensity of FAD and NAD(P)H
$$
R=\frac{\text{FAD}}{\text{FAD}+\text{NAD(P)H}}.
$$
For Fig. \ref{fig4}d, to visualize the 2D redox map, a binary mask was used to exclude the background region where both FAD and NAD(P)H had negligible signals. The multiphoton generation efficiency of FAD and NAD(P)H in the imaging system was calibrated using standard FAD (MilliporeSigma, F6625) and NAD(P)H (MilliporeSigma, N8129) solutions of 1\,mM.

\subsection*{Sample preparation for blood-brain barrier microfluidic model and IF staining}
The microvascular network of the blood-brain barrier was engineered employing a tri-culture of primary human astrocytes (ScienCell, 1800), primary human brain pericytes (ScienCell, 1200), and IPSCs-derived vascular endothelial cells (Alstem, iPS11). Cells were expanded using the following media: Astrocyte Medium (ScienCell, 1801), Pericyte Medium (ScienCell, 1201), VascuLife VEGF Endothelial Medium (Lifeline Cell Technology, LL-0003) supplemented with 10\% fetal bovine serum (Thermo Fisher Scientific, 26140-079) and SB 431542 (Selleckchem, S1067). Cells were passaged using TrypLE Express (ThermoFisher, 12604021). The 3D microfluidic co-culture was conducted as described in \cite{ko2023accelerating}.
For Fig. \ref{fig4}c, IF staining for anti-platelet-derived growth factor receptor beta (PDGFR-beta)  (Abcam, ab69506; RRID: AB1269704), and anti-glial fibrillary acidic protein (GFAP) (Abcam, ab10062; RRID: AB296804) was conducted as described in \cite{hajal2022engineered}.

\subsection*{Sample preparation for tracking monocyte behaviors \emph{in vitro}}
Human umbilical vein endothelial cells (HUVECs; Angioproteomie, cAP-0001) and normal human lung fibroblasts (NHLFs; Lonza, CC-2512) were cultured in VascuLife VEGF Endothelial Medium and FibroLife S2 Fibroblast Medium (Lifeline Cell Technology, LL-0011), respectively. After the cells were detached with TrypLE Express, they were suspended in a thrombin solution, mixed with fibrinogen, and seeded into a microfluidic device, as previously described \cite{zhang2022interstitial}. 10\,\textmu L of the cell-laden fibrin mix was seeded per device into a central channel that measured approximately 9\,mm$\times$3\,mm$\times$500\,\textmu m, with final cell concentrations of 7\,M/mL and 1.5\,M/mL for the HUVECs and NHLFs, respectively. Devices were kept in the incubator after gelation, with daily media changes of VascuLife medium for 6 days, until the HUVECs self-assembled into perfusable microvascular networks.
Primary human monocytes were isolated from buffy coats acquired from Massachusetts General Hospital. Monocytes were resuspended at 2\,M/mL before addition to microvascular networks.

% \begin{figure}[h]
% \centering
% \includegraphics[width=\textwidth]{figs/SLAM2_supp1.pdf}
% \caption{\textbf{Validation of 1100 nm 3PAF NAD(P)H imaging and 2PAF FAD imaging.} Reference images acquired from OPA at the same site using typical 2PAF NAD(P)H imaging excitation wavelength (750 nm (\textbf{a})) and 2PAF FAD imaging excitation wavelength (900 nm (\textbf{b})).}
% \label{validation}
% \end{figure}

% \begin{figure}[h]
% \centering
% \includegraphics[width=\textwidth]{figs/SLAM2_supp2.pdf}
% \caption{\textbf{Shorter fiber provides shorter band pulse duration.} \textbf{a}, Measuring the band pulse duration with 1100$\pm$25 nm bandpass filter. \textbf{b}, under the same input energy, the band pulse duration is shorter in a shorter fiber.}
% \label{validation}
% \end{figure}

\bibliographystyle{naturemag}

\bibliography{SLAM2}

\section*{Acknowledgement}
We thank Matthew Yeung, Steven F. Nagle, and Haidong Feng for helpful discussions.
We acknowledge that Fig. \ref{fig1}a and Fig. \ref{fig5}a were made in BioRender (https://biorender.com/).
\vspace{2mm}

\noindent\textbf{Funding:} The work has been supported by MIT startup funds and CZI Dynamic Imaging via Chan Zuckerberg Donor Advised Fund (DAF) through the Silicon Valley Community Foundation (SVCF).
K.L. acknowledges support from the MIT Irwin Mark Jacobs (1957) and Joan Klein Jacobs Presidential Fellowship.
H.C. acknowledges support from the MIT Kailath Fellowship.
K.S. acknowledges support from MIT Thomas and Sarah Kailath Fellowship.
F.M.P. was supported by the Postdoc. Mobility fellowship (P500PT 211085) from the Swiss National Science Foundation. 
\vspace{2mm}

\noindent\textbf{Author contributions:} K.L., H.C., T.Q., and S.Y. conceived the idea of the project. S.Y. supervised the research and obtained the funding. K.L., H.C., L.Y., and T.Q. built the optical setup. K.L. and K.S. performed the optical experiments and did the analysis. S.S., F.M.P., Z.W., E.L.K., E.N.T., M.L., E.L., R.D.K., L.G.G., and F.W. prepared and provided the samples. K.L., T.Q., and S.Y. wrote the manuscript with the input from all authors.
\vspace{2mm}

\noindent
\textbf{Competing interests:} S.Y., K.L., and H.C. are coinventors of a patent on the method and apparatus for deep tissue imaging. The other authors declare no competing interests.
\vspace{2mm}

\noindent
\textbf{Data and materials availability:} All data needed to evaluate the conclusions in the paper are present in the paper and/or the Supplementary Materials. Additional data related to this paper may be requested from the authors.
%
%\newpage
%\section*{Supplementary Material}
%
%\textbf{Supplementary Video 1} Imaging data for deep metabolic and structural imaging in the intact living human blood-brain barrier microfluidic model.
%\vspace{2mm}
%
%\noindent\textbf{Supplementary Video 2} Rendered 3D imaging for deep metabolic and structural imaging in the intact living human blood-brain barrier microfluidic model.
%\vspace{2mm}
%
%
%\noindent\textbf{Supplementary Video 3} Imaging data for dynamic metabolic and structural imaging of monocyte behaviors \textit{in vitro}.

\end{document}